\documentclass[manuscript]{aastex}

\usepackage{color}
\usepackage[normalem]{ulem}

\shorttitle{Galactic chemical evolution of $p$-nuclei}
\shortauthors{Travaglio et al.}

\begin{document}


\title{Testing the role of SNe Ia for Galactic chemical evolution of $p$-nuclei with 
2D models and with $s$-process seeds at different metallicities}

\author{C. Travaglio}
\affil{INAF - Astrophysical Observatory Turin, Strada Osservatorio 20, 10025 Pino Torinese (Turin), Italy}
\affil{B2FH Association - Turin, Italy}
\email{travaglio@oato.inaf.it, claudia.travaglio@b2fh.org}

\author{R. Gallino}
\affil{Dipartimento di Fisica, Universit\`a di Torino, Via P.Giuria 1, 10125 Turin, Italy}
\affil{B2FH Association - Turin, Italy}

\author{T. Rauscher}
\affil{Centre for Astrophysics Research, School of Physics, Astronomy and Mathematics,
University of Hertfordshire, Hatfield AL10 9AB, United Kingdom}
\affil{Department of Physics, University of Basel, 4056 Basel, Switzerland}
\affil{UK Network for Bridging Disciplines of Galactic Chemical Evolution
(BRIDGCE), \url{http://www.astro.keele.ac.uk/bridgce}, United Kingdom}

\author{F. K. R{\"o}pke}
\affil{Universit{\"a}t W{\"u}rzburg, Am Hubland, D-97074 W{\"u}rzburg, Germany}

\and 

\author{W. Hillebrandt}
\affil{Max-Planck-Institut f\"ur Astrophysik, Karl-Schwarzschild-Str.~1, D-85748 Garching bei M\"unchen, Germany}

\begin{abstract}

The bulk of $p$ isotopes is created in the 'gamma processes' mainly by
sequences of photodisintegrations and beta decays in explosive
conditions in Type Ia supernovae (SNIa) or in core collapse supernovae (ccSN). 
The contribution of different stellar sources to the observed distribution 
of $p$-nuclei in the Solar System is still under debate.
We explore single degenerate Type Ia supernovae in the framework of
two-dimensional SNIa delayed-detonation explosion models.  Travaglio
et al. (2011) (hereafter TRV11) discussed the sensitivity of
$p$-nuclei production to different SNIa models, i.e. delayed
detonations of different strength, deflagrations, and the dependence
on selected $s$-process seed distributions. Here we present a detailed
study of $p$-process nucleosynthesis occuring in SNIa with $s$-process
seeds at different metallicities. Based on the delayed-detonation
model DDT-a of TRV11, we analyze the dependence of $p$-nucleosynthesis
on the $s$-seed distribution obtained from different strengths of the
$^{13}$C-pocket. We also demonstrate that $^{208}$Pb-seed alone changes the $p$-nuclei 
production considerably.
The heavy-$s$ seeds (140 $\le A <$ 208) contribute with about 30-40\% to 
the total light-$p$ nuclei production
up to $^{132}$Ba (with the exception of $^{94}$Mo and $^{130}$Ba, to which
the heavy-$s$ seeds contribute with about 15\% only). Using a Galactic
chemical evolution code (see Travaglio et al.~2004) we study 
the contribution of SNIa to the solar stable $p$-nuclei.
We find that explosions of Chandrasekhar-mass single degenerate systems
produce a large amount of $p$-nuclei in our Galaxy, both in the range of light
($A \le$ 120) and heavy $p$-nuclei, at almost flat 
average production factors (within a factor of about 3). We discussed in 
details $p$-isotopes such as $^{94}$Mo with a behavior diverging from the average, 
which we attribute to uncertainties in the nuclear data or in SNIa modelling.

Li et al.~(2011) find that about 70\% of all SNeIa are normal events. If these are
explained in the framework of explosions of Chandrasekhar-mass white dwarfs 
resulting from the single-degenerate progenitor channel, we find that they are 
responsible for at least 50\% of the $p$-nuclei abundances in the Solar System.

\end{abstract}


\section{Introduction}

The origin of heavy nuclei was discussed by Cameron (1957), who
called 35 species {\it excluded isotopes}. Indeed they are outside of
both the $s$ and $r$ neutron capture paths, and they are typically 
10$-$1000 times less abundant than the
corresponding $s$- and/or $r$-isotopes in the Solar System.  The
origin of $p$-nuclei was investigated starting with the pioneering
work of Cameron~(1957) and Burbidge et al.~(1957), and later by
Audouze \& Truran~(1975) and Arnould~(1976). The first work analyzing
the possibility of having efficient photodisintegrations in
Chandrasekhar-mass SNIa explosions was published by Howard, Meyer \&
Woosley~(1991). The initial $s$-seed distribution they used was
derived from helium flashes as calculated by Howard et al.~(1986). In
Figure~2 of that work, the authors claim that they can reproduce the
abundance pattern of all $p$-nuclei, including light-$p$ nuclei,
within a factor of about three. However, they obtained an overproduction
of $^{74}$Se, $^{78}$Kr, and $^{84}$Sr. The $p$-process abundances of
these three isotopes are very sensitive to the proton density, which
the authors considered rather uncertain. They also obtained a rather
low production of $^{94}$Mo and $^{96}$Ru with respect to the other
light $p$-nuclei. A detailed discussion on these light $p$-nuclei will
be given in Section~4 of the present paper. Later Goriely et al.~2002,
Goriely et al.~2005, and Arnould \& Goriely~2006, analyzed the
$p$-process production in He-detonation models for sub-Chandrasekhar
mass WDs. These authors considered as seeds $s$-process solar
abundances. They found Ca to Fe to be overabundant with respect to
$p$-nuclei (with the exception of $^{78}$Kr) by a factor of
$\simeq$100. They concluded that a He detonation is not an efficient
scenario for the production of the bulk Solar-System $p$-isotopes.
Kusakabe et al.~(2011) presented $p$-process nucleosynthesis
calculations in a CO-deflagration model of SNIa, i.e., the W7 model of
Nomoto et al.~(1984).  Similar to Howard et al.~(1986), they assumed
enhanced $s$-seed distributions using the classical $s$-process
analysis, testing two different mean neutron exposures $\tau_o$, a
flat distribution for $\tau_o$ = 0.30 mb$^{-1}$, and a decreasing
$s$-process distribution corresponding to $\tau_o$ = 0.15 mb$^{-1}$.

They noticed that for a flat $s$-seed enhanced distribution the
production factors of light $p$-nuclei show a strong deficiency in
the range $^{78}$Se to $^{98}$Ru. From this, they concluded that 
SNIa may have contributed to the enrichment of $p$-nuclei more 
effectively than ccSNe.

In our previous TRV11 paper, we obtained a consistent production of
$p$-nuclei in the single-degenerate Chandrasekhar mass WD explosion scenario. We
presented two-dimensional hydrodynamic models of SNIa. The corresponding 
nucleosynthesis was calculated in a post-processing step following the thermal history of
Lagrangian tracer particles. We found a significant production of
$p$-nuclei from these stellar explosions, at the same level compared to $^{56}$Fe for light
as well as heavy $p$-nuclei. We demonstrated that our model is able to produce
light and heavy $p$-nuclei in one single process.  In our analysis, 
we assumed enhanced $s$-seed distributions directly obtained from a
sequence of thermal pulses in the material accreted onto the
  exploding white dwarf from a normal companion star (see Gallino et
al. 1998). In the context of light-$p$ nuclei production, the major
problem discussed in TRV11 is the resulting abundance of $^{94}$Mo, found to be far too low relative
to the abundances of the other light-$p$ nuclei.  We also found an important
contribution from $p$-process nucleosynthesis to $^{80}$Kr and
$^{86}$Sr (originally considered $s$-only nuclei), to the neutron magic
$^{90}$Zr, and to the neutron-rich $^{96}$Zr (due to neutron captures
from the residual abundance of $^{22}$Ne during the explosive phase).
Concerning the heavy $p$-isotopes, the $s$-process nature of
$^{152}$Gd has been later confirmed by different works on $s$-process
nucleosynthesis (Gallino et al. 1998; Bisterzo et al. 2010), where a
predominant $s$-process origin was demonstrated (see also discussion
in TRV11).

In the present paper we investigate for the first time in the literature
the effect of metallicity on $p$-process nucleosynthesis in SNIa,
starting with a range of $s$-seed distributions obtained for different
metallicities. Using a simple chemical evolution code (Travaglio et
al.~2004), we estimate the contribution of SNIa to the solar
$p$-process composition. The same study was recently done for
radiogenic $p$-isotopes by Travaglio et al.~(2014).  

It is currently impossible to discuss the interplay between the role of 
SNeIa and ccSNe in the production of $p$-nuclei in the Galaxy. 
Infact a complete study of $p$-nucleosynthesis with metallicity is missed for
ccSNe. But also from the observational point of view, unfortunatey there is no way to measure 
observed chemical evolution of $p$-nuclei since they are too rare with respect 
to the $s$- and $r$-fractions. For example, there are recent attempts to observe elements like Mo and Ru (very 
debated for $p$-process nucleosynthesis studies) in field stars of the Galaxy at different metallicities 
(see e.g. Hansen et al. 2014; Peterson 2013). Only a small fraction of the elements 
Mo and Ru are $p$: 14.84\% of the total Mo is $^{92}$Mo and only 9.25\% is $^{94}$Mo; for Ru, only 5.5\% of the element is $^{96}$Ru
and 1.88\% is $^{98}$Ru.
The $p$-fractions indicated above are therefore too small to be of any interest when the elements are observed in field stars. Only when will be possible to 
observe  isotopes of Mo and Ru in the spectra then we will be able to give interesting indications on the details of their nucleosynthesis processes.

For this work, the adopted SNIa model (described in detail in TRV11) is summarized in Section~2,
together with a brief description of the tracer particles method used
for nucleosynthesis calculations. In Section~3 the $s$-process seed
distributions considered in our study are described in detail. The
resulting $p$-process production and the effect of metallicity on
$p$-nuclei is discussed in Section~4, where also an analysis of the
various production mechanisms of the $p$-nuclei will be performed,
taking into account nuclear uncertainties. In Section~5 we analyzed in
detail how $s$-seeds with
different atomic mass number contribute to the $p$-nuclei production. Galactic
chemical evolution calculations are presented in Section~6, and the contribution
of SNIa to the solar system $p$-nuclei abundances is studied. In
Section~7 we discuss the contribution of SNIa to the production of
radiogenic $^{92}$Nd, $^{146}$Sm and $^{97,98}$Tc. Finally,
conclusions and work in progress are drawn in Section~8.

\section{Type Ia supernova models and tracer particles}

We used the SNIa-explosion model (DDT-a) described in detail by
TRV11. It is a representative example of the single-degenerate
scenario in which the WD has accreted material from a main-sequence or
evolved companion star until it finally approaches the Chandrasekhar mass and explodes as a delayed detonation. 
The model is based on the two-dimensional simulations presented by Kasen et al 2009.

The explosion itself is simulated in 2D by means of the combustion code LEAFS 
(Reinecke et al.~1999; Reinecke et al.~2002; R\"opke~2005; R\"opke \& Hillebrandt~2005)
which follows the evolution with an Eulerian grid. In order to compute 
for each zone of the star the history of temperature and density over 
time for each zone of the star we introduce a Lagrangian component in the form of tracer
particles. With the tracer-particle method it is possible to
reconstruct the ensuing nucleosynthesis. The nuclear post-processing
calculations are performed separately for each tracer. Summing the
chemical composition over all tracer particles gives the total
yields.  The tracer particles method was first introduced by Nagataki
et al.~(1997) for ccSSNe, and by Travaglio et al.~(2004b,
2005) for SNIa.

For 2D simulations it has been verified that 51,200 particles as used
here, uniformly distributed in mass coordinates, give sufficient
resolution (Seitenzahl et al.~2010). For each tracer particle we follow
the explosive nucleosynthesis with a detailed nuclear reaction network
for all isotopes up to $^{209}$Bi.  We select tracers within the
typical temperature range for $p$-process production,
i.e. $(1.5-3.7)\times 10^{9}$ K, and analyze their behavior in detail,
exploring the influence of different $s$-process seeds on the
$p$-process nucleosynthesis. In order to determine
the $s$-process enrichment prior to the explosion, we assume recurrent 
flashes occurring in the He-shell during the accretion phase with
neutrons mainly released by the $^{13}$C($\alpha$,n)$^{16}$O reaction
(Iben 1981; TRV11). This applies under the assumption that a small
amount of protons are ingested in the top layers of the He
intershell. Protons are captured by the abundant $^{12}$C and
convert it into $^{13}$C via
$^{12}$C($p$,$\gamma$)$^{13}$N($\beta^+\nu$)$^{13}$C at $T \simeq
1\times 10^8$ K.

\section{$s$-seeds at different metallicities}

In our model, $p$-process nucleosynthesis occurs in SNIa starting from a
pre-explosion $s$-process enriched seed composition. Therefore, it is
essential to determine the $s$-process enrichment prior to the
explosion.  Here, we assume enhanced $s$-distributions produced directly by 
a sequence of thermal-pulse instabilities in the accreted
material.  This idea has been described in detail by TRV11 and was
previously discussed by Iben~(1981), Iben \& Tutukov~(1991), and
Howard \& Meyer~(1993).

To be more specific, we assume recurrent flashes to occur in the
He-shell during the accretion phase. The matter accumulated onto the
carbon-oxygen white dwarf (hereafter CO-WD) therefore becomes enriched 
in $s$-nuclei. However, the mass
involved and the physical properties of the $^{13}$C-pocket, providing 
the free neutrons for the $s$-process, still have to be
considered as free parameters. Since no physical models are available we
explore different $s$-process distributions in order to better understand
the dependence of our results on these initial seeds (see also the
discussion in TRV11 and Travaglio et al.~2012).

For the present work we calculate $s$-process distributions for
8 metallicities, i.e. $Z =$0.02, 0.019, 0.015, 0.012, 0.011,
0.010, 0.006, 0.003 (a refined $s$-seed metallicity grid is necessary for chemical evolution calculations 
since the $s$-seeds are strongly dependent on metallicity), and 
we interpolate for all the other metallicities in between in order to calculate 
Galactic chemical evolution. We also investigate the effect of the $s$-seeds 
with  different $^{13}$C-pocket properties
(ST$\times$2, ST$\times$1.3, ST, ST/1.5, where $\sim$4 $\times$
10$^{-6} M_\odot$ of $^{13}$C in the pocket corresponds to the ST
case, Gallino et al.~1998; see Gallino et al.~1998 and Bisterzo et
al.~2010 for a detailed discussion of $^{13}$C-pocket profiles). The
$s$-process seeds adopted are shown in Figure~1 for the ST$\times$2
$^{13}$C-pocket case, and metallicites of $Z =$0.01, 0.006,
0.004, 0.003. In Figure~2 we show the average of four $^{13}$C-pockets
at different metallicities. In all figures the s-seed abundances are
normalized to the solar abundances of Lodders~(2009).

As discussed by Gallino et al.~(1998) and Travaglio et al.~(1999), the
synthesis of heavy nuclei requires neutron captures starting from Fe
seeds, so that the $s$-process is expected to decrease with decreasing
metallicity, i.e. to be of {\it secondary} nature. However, the
abundances produced depend not only on the initial Fe concentration
but also on the neutron exposure. The concentration of $^{13}$C in the
pocket is of {\it primary-like} nature (it is built from H and freshly
made C, and hence is independent of metallicity), while the abundance
of the neutron absorber $^{56}$Fe varies linearly with $Z$.
Therefore, for a given amount of $^{13}$C in the pocket, the neutron
exposure (proportional to the ratio $^{13}$C/$^{56}$Fe) is expected to
increase linearly with decreasing metallicity. This dependence would
compensate for the secondary nature of the $s$-elements, if the yields
of $s$-nuclei were linearly dependent on both neutron exposure and
$Z$. In addition, the behavior of $^{208}$Pb has to be considered
separately (Travaglio et al.~2001).  The gradual increase of the
neutron exposure towards low metallicities masks the expected secondary
behavior (see also Clayton~1988), resulting in a rather complex
dependence of $s$-process yields on metallicity. For the lower metal
content the neutron flux feeds Pb (in particular $^{208}$Pb).  A
clear understanding of the $s$-seeds behavior versus metallicity,
including the production of neutron-magic $^{208}$Pb at the termination
of the $s$-process path, is very important for the nature of
$p$-process (see Section~4).

Figure~1 shows a variation by a factor of $\simeq$10 for the 
abundances of the light $s$-only isotopes (up to $A\simeq$140) 
when $Z =$ varies from 0.003 to 0.01, and a spread of $\simeq$5 for heavy
$s$-only nuclei (with $A >$140). For $^{208}$Pb, the variation is
by a factor of $\simeq$8, but its trend is inverted with respect the 
behavior of other $s$-only
isotopes (i.e. lower $Z$ gives higher $^{208}$Pb abundance). In
Figure~2 we plot the average of four different $^{13}$C-pockets
(ST$\times$2, ST$\times$1.3, ST, ST/1.5) for the metallicities
discussed above.  The behavior of the $s$-only isotopes is different from 
that seen in Figure~1: still a rough dependence on $Z$ for light $s$-nuclei in the range 90$<
A <$ 120 is observed, but the variation is now reduced by a factor of 3. 
The spread is progressively decreasing for
higher metallicities. Further on, for heavy $s$-isotopes with 140 $< A
\le$ 204, an almost unique and flat distribution is obtained,
independent of metallicity with an enhancement around 2000 times
solar. Eventually, for $^{208}$Pb a spread of about a factor of 4
stands out again, with the lowest metallicity showing the highest
$^{208}$Pb abundance, at the level of 5000$\times$ solar.  The general
trend sketched above can be understood in the light of the typical
$s$-process enhancement occurring in AGB stars. For a given
$^{13}$C-pocket strength, decreasing the metallicity the $s$-flow
feeds more and more $^{208}$Pb at the termination of the $s$-path. At
the same time a progressive depletion of the lighter $s$-process
isotopes is found up to the magic neutron number $N =$ 82 ($^{138}$Ba
to $^{142}$Nd). In the region between the magic neutron number nuclei $N=$82 and
$N=$126, an almost flat $s$-process production factor ensues. A
similar trend occurs at a fixed metallicity by increasing the
$^{13}$C-pocket strength. We recall here that case STx2 is around the
maximum $^{13}$C-pocket strength we can reach, beyond which further
proton ingestion during a third dredge up episode from the envelope
would result in a decrease of $^{13}$C and production of the neutron
poison $^{14}$N (see the review by Busso, Gallino, \&
Wasserburg~1999).  In other words, with decreasing metallicity the
otherwise flat $s$-process distribution near the neutron magic numbers
N = 50 amd N = 82, corresponding to atomic mass numbers around  
$A=$90 and $A=$140, and also at the termination of the $s$-path at
around $A=$208 are progressively distorted. 

In Figure~3 we show the $s$-seeds for the range of $^{13}$C-pockets
(ST$\times$2, ST$\times$1.3, ST, ST/1.5, ST/2) and metallicities ($Z
=$ 0.02, 0.019, 0.015, 0.012, 0.011, 0.010, 0.006, 0.003) we cover in the
Galactic chemical evolution calculations (see Section~6 for discussion).

\section{$p$-process at different metallicities}

The $p$-process nucleosynthesis is calculated using a nuclear network
with 1024 species from neutrons and protons up to $^{209}$Bi combined
with neutron, proton, and $\alpha$-induced reactions and their
inverse. The code used for this work was originally developed and
presented by Thielemann et al.~(1996).  We employ the nuclear reaction
rates based on the experimental values and the Hauser \& Feshbach
statistical model NON-SMOKER (Rauscher \& Thielemann~2000), including
the recent experimental results of Maxwellian-averaged neutron capture
cross section of various $p$-only isotopes (Dillmann et al.~2010;
Marganiec et al.~2010).  Theoretical and experimental electron capture
and $\beta$-decay rates are from Langanke \& Mart\'inez-Pinedo~(2000).

We discuss in this Section the sensitivity of $p$-process production
to $s$-seeds at different metallicites. We analyze the 
{\it primary}/{\it secondary} nature of the resulting
$p$-nuclei. According to Rauscher et al.~(2013) (and references
therein) the $p$-process is of {\it secondary} nature and scales with
the amount of seed nuclei in the star. In this work we present our
results obtained with SNIa models which only partly confirm this
statement.

In Figure~4 we plot the resulting $p$-process abundances, starting
from $^{74}$Se, obtained by using different $s$-seeds at different
metallicities. On the axis of ordinates the production factor of each isotope is
plotted with respect to solar, normalized to $^{56}$Fe. Note that the
abundances of $p$-nuclei heavier than $A =$ 100 are much higher than a
factor of $\sim$3 times their solar value. However, in this Figure
we plot the nucleosynthesis resulting from one single star and not
the integrated abundances over all the Galaxy (see Section~6 for
discussion).  The choice for these $^{13}$C-pockets and metallicities
will be used for our best fit of Galactic chemical evolution
calculation.

In order to better understand the dependence of $p$-nuclei production
on metallicity and the $^{13}$C-pocket, we show in two separate
figures (Figure~5 and Figure~6) the behavior of $p$-process
nucleosynthesis as a function of metallicity and the $^{13}$C-pocket,
respectively.  From Figure~5 we can see that the first three $p$-only
isotopes, $^{74}$Se, $^{78}$Kr, and $^{84}$Sr, show a secondary
behavior: their abundances scale almost linearly with $Z$. They depend
mainly (for about 60-70\%) on the light (up to $A\simeq$140) $s$-seeds 
that, as shown in Figure~1 and Figure~2, are strongly
dependent on metallicity.  Starting from $^{90}$Zr (even if this
isotope is not $p$-only it has to be included in the discussion, see
also TRV11) and $^{92}$Mo and up to $^{138}$Ba, the $p$-nuclei show a
very weak dependence on $Z$. The isotopes in this atomic mass-number
range are mainly produced by photodisintegration from the heavy
$s$-seeds isotopes, thus showing a primary-like behavior (see
Figure~6). Isotopes in the region from $^{136}$Ce up to $^{196}$Hg
scale with metallicity but show the opposite trend, i.e. highest
abundances for the lowest metallicities. TRV11 (see also Dillmann et
al.~2008a) discussed the fact that Pb-seeds are converted to nuclei of
lower mass by photodisintegration sequences starting with ($\gamma$,n)
reactions. Therefore, an important contribution to the heavy $p$-only
isotopes is obtained.

We also found, as discussed in TRV11, that the isotopes $^{113}$In,
$^{115}$Sn, $^{138}$La, $^{152}$Gd, and $^{180m}$Ta, diverge from the
average $p$-process production. Among them, $^{152}$Gd and $^{180m}$Ta
have an important contribution from the $s$-process in AGB stars
(Arlandini et al. 1999), or the neutrino process in ccSN (Woosley et
al.~1990; Wanajo et al.~2011). Both $^{113}$In and $^{115}$Sn are not
fed by the $p$-process nor by the $s$-process. For these, we refer to
the discussion of Dillmann et al. (2008b) and TRV11.

The still puzzling $^{94}$Mo deserves special attention. Can
nuclear uncertainties account for the low $^{94}$Mo yield, compared to
the other $p$ nuclei, in our models? Howard et al.~(1991) demonstrated
that most of the $^{94}$Mo is produced from $^{98}$Mo through a
($\gamma$,n) chain.

We confirmed that a similar reaction chain also acts in our
models. To illustrate this, Figure\ \ref{fig:flowTracers} shows three
exemplary trajectories for which time-integrated reaction flows are
plotted in Figures\ \ref{fig:flow94Momax}$-$\ref{fig:flow94Momin}. The
flow shown in Figure\ \ref{fig:flow94Momax} gives a maximum production
of $^{94}$Mo (the corresponding trajectory is labeled
``$^{94}$Mo$_\mathrm{max}$'' in Figure\ \ref{fig:flowTracers}).  The one
shown in Figure\ \ref{fig:flow94Modrop} also favors $^{94}$Mo production but
at a lower level (the corresponding trajectory is labeled
``$^{94}$Mo$_\mathrm{drop}$'' in Figure\ \ref{fig:flowTracers}). In both
cases, $^{94}$Mo is fed through ($\gamma$,n) reaction sequences, with
the strongest flow originating at $^{98}$Mo, which,in turn, is replenished 
to some extent also by a small flow originating from $^{100}$Mo. It should be
noted that the reactivities for this ($\gamma$,n) sequence are
experimentally well constrained, as they involve stable nuclei and the
corresponding (n,$\gamma$) reaction cross sections have been measured
(Dillmann et al.\ 2006). Thermal population of excited states does not
play a role in these nuclei and thus the measured (n,$\gamma$) cross
sections allow to compute the (n,$\gamma$) and ($\gamma$,n) rates
without further theory uncertainties (Rauscher 2012, 2014). Therefore
this part of the flow does not bear large nuclear uncertainties, it is
rather determined by the seed abundances received by the stable Mo
isotopes. Some production of $^{94}$Mo is also found via
$^{95}$Tc($\gamma$,p)$^{94}$Mo. Its contribution to $^{94}$Mo,
however, is an order of magnitude lower than that of the ($\gamma$,n)
sequence and thus it does not contribute appreciably to the
uncertainty although the $^{94}$Mo(p,$\gamma$) reactivities are
unmeasured.

Consequently, the only significant nuclear uncertainty is found in the
destruction of $^{94}$Mo by $^{94}$Mo($\gamma$,n)$^{93}$Mo. As already
discussed by Travaglio et al.~(2014), the sequence
$^{94}$Mo($\gamma$,n)$^{93}$Mo($\gamma$,n)$^{92}$Mo, leads to
production of $^{92}$Mo, with $^{94}$Mo($\gamma$,n)$^{93}$Mo being the
slower, and thus determining, reaction here. A conservative estimate
for the uncertainty of this rate (taken from Rauscher \& Thielemann
2000) is a factor of 2. 

The comparison between the ``$^{94}$Mo$_\mathrm{max}$'' and
``$^{94}$Mo$_\mathrm{drop}$'' trajectories in
Figure\ \ref{fig:flowTracers} shows that ``$^{94}$Mo$_\mathrm{drop}$''
reaches a slightly higher peak temperature. This higher temperature
not only increases the ($\gamma$,n) flow, that destroys $^{94}$Mo without
increasing the production notably but also enables further
destruction of $^{94}$Mo by $^{94}$Mo($\gamma$,$\alpha$)$^{90}$Zr. The
latter reaction is offset by a slight enhancement in
$^{95}$Tc($\gamma$,p)$^{94}$Mo but the flow through this reaction is
still lower by a factor of 0.1 than the increased destruction
flow. This illustrates that $^{94}$Mo can only be produced within a
narrow temperature window: at low temperature, the ($\gamma$,n) flow
is small or non-existent, at too high temperature $^{94}$Mo is
destroyed by additional reactions. Prerequisite for the efficient
production of $^{94}$Mo by photodisintegration in any site is that the
mainly contributing trajectories spend as much time as possible in
this temperature window.

Finally, Figure\ \ref{fig:flow94Momin} shows the flows for a trajectory
leading to a minimal production of $^{94}$Mo although the achieved
peak temperature is similar to that in the
``$^{94}$Mo$_\mathrm{max}$'' case (the corresponding trajectory is
labeled ``$^{94}$Mo$_\mathrm{min}$'' in Figure\ \ref{fig:flowTracers};
incidentally this is the same trajectory giving a maximum $^{92}$Nb
production as shown in Figure\ 5 of Travaglio et al.~2014). Inspection
of the flows shows that all ($\gamma$,n) flows in this region are
significantly suppressed compared to the ``$^{94}$Mo$_\mathrm{max}$''
case in Figure\ \ref{fig:flow94Momax}. The $^{94}$Mo$\longrightarrow
^{93}$Mo flow is too small to show in the plotted range. The
$^{91}$Zr$\longrightarrow ^{92}$Zr flow is even replaced by its
reverse. The key to understanding the difference is the fact that the
``$^{94}$Mo$_\mathrm{min}$'' trajectory reaches photodisintegration
temperatures only at much higher density (note the logarithmic scale
of the horizontal axis in Figure\ \ref{fig:flowTracers}). At all times,
$A$(n,$\gamma$)$B$ and $B$($\gamma$,n)$A$ rates are competing. Their
relative strengths are on one hand determined by the reaction
$Q$-value (which is given through the well-known nuclear masses)
but on the other hand (n,$\gamma$) rates also scale with the available
neutron density whereas ($\gamma$,n) do not (Rauscher~2011). At the
temperatures at which photodisintegration of nuclei in the Mo region
becomes possible, heavier nuclei are already significantly destroyed
because they are less tightly bound (their $Q$-value for (n,$\gamma$)
is lower than that of lighter nuclei; see also Rauscher et
al.~2013). This allows (n,$\gamma$) reactions to occur in the lighter
region and at high density they will be faster than their ($\gamma$,n)
counterparts. This illustrates the important point that significant
production of $p$-nuclei in a $\gamma$-process is possible only when
the densities remain limited, i.e. it depends sensitively on
the thermodynamic histories of the explosive layers. For example, a
model in which more trajectories experience lower densities during
$\gamma$-processing would lead to increased $^{94}$Mo production. 
However, since this scenario will affect all (n,$\gamma$)/($\gamma$,n) ratios
and thus it is unclear whether it would lead to an enhancement of
the final $^{94}$Mo with respect to the other light $p$-nuclides. This
has to be investigated in detail in future calculations because the
impact of enhanced (n,$\gamma$) rates at higher density is not trivial
as it depends on the specific $Q$-values in the ($\gamma$,n) chains
producing specific isotopes. Furthermore, it will also impact other
reaction types (such as (p,n), (p,$\gamma$), and their reverses) in
different ways and thus the final outcome strongly depends on the
actual reaction sequences producing and destroying a specific
nucleus. This may change the ratio of the $^{94}$Mo abundance also
relative to abundances of other $p$-nuclei but it remains an open
question whether the required relative increase by a factor of 
10 can be achieved.

\section{The role of $s$-seeds of different atomic mass number for
  $p$-process nucleosynthesis}

In order to understand in detail the $s$-seed origin of each
$p$-nucleus, we performed the following study: fixing the metallicity
($Z =$ 0.006, i.e., to a value where we find the highest production of
$p$-nuclei in our Galactic chemical evolution calculations, see Section~6) 
and fixing a $^{13}$C-pocket (i.e., STx2 which is the highest value we used), we
first tested the role of the $s$-seed $^{208}$Pb alone, by assigning solar abundances 
to all other $s$-seed nuclei. Details of the
resulting effect on $p$-nucleosynthesis are given in Table
\ref{tab:z0p006} (third column for the $^{208}$Pb only case, and
second column for the standard case). A $^{208}$Pb seed alone contributes
about 30\% to $^{74}$Se, and 10-20\% (or less) to all the other
light-$p$ nuclei up to $^{158}$Dy (with the exception of $^{144}$Sm,
where $^{208}$Pb alone accounts for about 45\%). The contribution to the 
isotopes between $^{162}$Er and $^{190}$Pt from $^{208}$Pb is about 40\%. 
Finally, for the heaviest of the $p$-only nuclei, $^{196}$Hg, we find the 
highest share (~60\%) in production originating from $^{208}$Pb.  

In Table \ref{tab:snuc} we list all $s$-isotopes which we find to have
$p$-contribution. For these isotopes we report in the last column
the effect of $^{208}$Pb. We find that most of these
isotopes (from $^{80}$Kr up to $^{176}$Hf), the $^{208}$Pb seed
contributes about 10-20\% (or less).  A similar analysis has been
carried out accounting for either as $s$-nuclei either the only heavy $s$-nuclei (140
$\le A <$ 208) or the light $s$-nuclei (70 $\le A <$ 140) only.  The
results are shown in Table \ref{tab:z0p006}: in Column 6 the results
for the heavy $s$-seed only case and in Column 7 the light $s$-seed
only case are given. Not surprisingly, the heavy $s$-nuclei are most important 
for the heavy $p$-nuclei, in particular for
$^{184}$Os (100\% of contribution), $^{158}$Dy (89\%), 
$^{136,138}$Ce (about 84\% and 75\%, respectively),
$^{190}$Pt (about 73\%), and $^{156}$Dy (about 70\%), but there are also
significant contributions of about 40-60\% are to the other heavy $p$-isotopes in
the region from $^{144}$Sm up to $^{196}$Hg. We also find $\sim$30-40\% 
of the light $p$-nuclei up to $^{132}$Ba to originate from heavy $s$-seeds, 
with the exception of $^{94}$Mo and $^{130}$Ba, which have a lower 
share of $\sim$15\%. In
contrast, light $s$-seeds are the main producers of the
light $p$-isotopes, and they account for about 73\% for $^{74}$Se
and $^{94}$Mo (where important contribution from other stellar sources
or errors in reaction rates can be estimated), and finally $^{130}$Ba
(about 80\%). All other light-$p$ isotopes are produced from
light-$s$ seeds with typical sharesin the rage from 40 to
60\%. The heaviest $p$-nuclei are almost unaffected by the light-$s$
seeds.

In the fourth columns of Table \ref{tab:z0p006} and Table \ref{tab:snuc} we show 
results of a second test for comparison, where we retain a metallicity of 
$Z =$ 0.006, but reduce the strenght of the $^{13}$C-pocket by using 
the STx1.3 model. For this case we test the
role of $^{208}$Pb seed for $p$-process nucleosynthesis, as we did
in the above standard case (see fifth column in Tables \ref{tab:z0p006},
\ref{tab:snuc}). Comparing the two cases presented in Table 1 (column 3 and 5), we notice that 
typically the case with STx2 $^{13}$C-pocket and a $^{208}$Pb seed
contributes by a factor of $\sim$2 to the light-$p$ nuclei
with respect to the case STx1.3 $^{13}$C-pocket and $^{208}$Pb seed. 
For the abundance of $^{144}$Sm we obtain identical results for 
STx2 and STx1.3 cases. In contrast, however, we observe an increased production of
the heavy-$p$ nuclei for the case STx2 and $^{208}$Pb seed by $\sim$10\%. 
The only exception is for the heaviest $p$-nucleus $^{196}$Hg, where we find a highest abundance 
in the case STx1.3 $^{208}$Pb alone.  The reason for this
behavior is that at the same metallicity ($Z =$ 0.006)
the $^{13}$C-pocket STx1.3 is less efficient than the $^{13}$C-pocket
STx2 for the production of heavy-$s$ as well as for $^{208}$Pb.

\section{Galactic chemical evolution}

The main goal of this work is to provide predictions for
Galactic chemical evolution of $p$-nuclei. For this, we employ  
the Galactic chemical evolution code presented by Travaglio
et al.~1999, 2001, 2004. The model considers the
Galaxy as the evolution of three interconnected zones, halo, thick
disk and thin disk. The matrix of the isotopes within the chemical
evolution code was set to cover all the light nuclei up to the
Fe-group, and all the heavy nuclei along the $s$-process path up to
$^{209}$Bi. For the present work we extended the matrix of the
isotopes to account for the $p$-nuclei and we followed their evolution
over time/metallicity until solar metallicity was reached. We included
in the code the $p$-nuclei abundances obtained from our SNIa model at various metallicities
as discussed in Section~4, and interpolate between them smoothly. 
In Figure~7 we show the resulting $p$-process production factors
taken at the epoch of Solar System formation for nuclei in the atomic
mass number range 70 $\le A \le$ 210. To be more clear we note that for these results we 
only include SNIa for the contribution to $p$-nuclei.
Our choice for the $s$-seeds was introduced in Section~3 (see Figure~3). As discussed in the previous
section, it is clear that a few nuclei originally ascribed to the
$p$-only group ($^{113}$In, $^{115}$Sn, $^{138}$La, $^{152}$Gd, and
$^{180m}$Ta) are far below the average of the other $p$-nuclei. Thus, 
if this model is correct, they should be ascribed to
different astrophysical sources.
As shown in Figure 7 and detailed in Table 3, we find in our Galactic chemical 
evolution calculations that all the other 
$p$-only isotopes are---within a factor of about three---produced at the Solar System composition.
The lightest $p$-nucleus $^{78}$Kr and the heaviest $^{158}$Dy and $^{180}$W can be ascribed to
the same production site when an additional uncertainty factor of two is included 
(which seems reasonable given the large uncertainties in photodisintegration rates).
Therefore, for the first time, we are able to explain the synthesis of almost all $p$-isotopes production 
in one single scenario.  The most striking problem we face is the very low relative abundance of
the true $p$-only $^{94}$Mo with respect the the average of all other $p$-only
nuclei, which is possibly related to the theoretical estimate of the neutron capture on
unstable $^{93}$Mo. The effect of the nuclear uncertainties will be explored in a forthcoming paper. 

Figure 7 and Table 4 also show that neutron magic $^{90}$Zr and especially 
$^{96}$Zr receive an important contribution by the $p$-process (about 20\% and 40\%,
respectively). Moreover, it is not excluded that the most proton-rich $s$-only isotopes 
for a given element may receive some contribution from the $p$-process.  As illustrated
in Figure 7, this is the case for $^{80}$Kr and $^{86}$Sr, with $p$-contribution of the
order of 10\%. However, before examining the problem in more detail, an analysis of the 
uncertainties in the involved reaction rates is necessary. Out of the light $p$-nuclei the isotopes $^{78}$Kr, 
$^{94}$Mo, $^{108}$Cd, and $^{114}$Sn are by a factor of three or more (for $^{94}$Mo) less abundant 
compared to the solar value. Among the heavy-$p$ nuclei $^{158}$Dy is also by more than a factor of 
three below the solar abundance. In contrast, we found $^{180}$W higher by a factor of more than three 
with respect to solar. Under the hypothesis 
that SNIa are responsible for 2/3 of solar $^{56}$Fe, and assuming that
our DDT-a model represents the typical SNIa with a frequency of 70\% (Li
et al. 2011), we conclude that they are responsible for at least 50\% of
all $p$-nuclei.

With the same approach, Travaglio et al.~(2014) discussed the origin
of the short-lived radionuclides $^{92}$Nb and $^{146}$Sm. They compared the
value of the ratio between the abundances of these radionuclides and those 
of the corresponding stable reference isotopes $^{92}$Mo and
$^{144}$Sm to what has been recently measured in meteorites (Rauscher
et al.~2013). The conclusion is that SNIa can also play a key role in
the production of $^{92}$Nb and $^{146}$Sm, but nuclear uncertainties 
have to be taken into account.

In the framework of Galactic chemical evolution of $p$-nuclei,
the role of ccSN has also to be taken into account. Rauscher et
al.~(2002) followed the $\gamma$-process through the presupernova
stages and the supernova explosion. As expected for the weak
$s$-process component in massive stars, only $s$-nuclei in the mass 
range $64\le A \le 88$ are produced and occurred in
situ prior to the explosion phase. Rauscher et al.~(2002) presented 
results for 15, 19, 21, 25 $M_\odot$ ccSNe modeled in spherical symmetry
and with initial solar metallicity. For the 15, 21, and 25
$M_\odot$ models, proton-rich heavy isotopes in the mass ranges
$124\le A \le 150$ and $168\le A \le 200$ were produced in solar
abundance ratios within about a factor of two relative to $^{16}$O, the
most abundant nucleus in the ejecta of ccSNe. For mass numbers $A \leq 124$ and
$150\le A \le 165$ the production of the $p$-isotopes is down by about
a factor of $3-4$. While the main $\gamma$-process synthesizes
$p$-nuclei through photodisintegration reactions during the SN
shockfront passage, some of the models showed pre-explosive
$p$-production due to a high entropy in the O/Ne shell of the evolved
star. Most of this is wiped out again when the supernova shock sweeps
through the layer. Nevertheless, depending on the adopted convection
model (see also Bazan \& Arnett~1994), some light, strongly bound
$p$-nuclei may survive from pre-explosive production. This behavior
complicates predictions for the contribution of ccSNe to the solar
composition of $p$-nuclei. In addition, it has to be taken into account 
that the SNIa scenario discussed here may not explain all normal events.
Thus, alternative scenarios for $p$-process nucleosynthesis in binary systems 
should be explored, such as SNeIa from WD-WD mergers (e.g. Pakmor et al.~2010, 2012) or 
double detonations in sub-Chandrasekhar mass WDs (e.g. Fink et al.~2010).

\section{Conclusions}

We have presented results of detailed $p$-process nucleosynthesis
calculations for two-dimensional models of delayed detonations
in Chandrasekhar-mass WDs resulting from the single degenerate 
progenitor scenario. In these SNIa models, the nucleosynthesis 
was followed by the tracer-particles method. 
The initial $s$-seeds were assumed to be
created during the mass accretion phase. Since up to now no
nucleosynthesis calculations of the accretion phase are available our
hypothesis is based on the assumption that a small amount of protons
are ingested at the top layers of the He intershell. Following the
work of TRV11 on the nucleosynthesis for a solar
metallicity SNIa model in the same scenario, we tested and discussed 
the consequences of different amounts of $^{13}$C and different
metallicites on the synthesis of $p$-nuclei. We demonstrated that the
$^{208}$Pb $s$-seed alone plays an important role for $p$-nuclei
production, due to photodisintegration chains starting from the
heaviest nuclei and going down in mass number. We analyzed the
dependence of all the $p$-isotopes on metallicity, and we identified
the isotopes with a weak (like $^{92}$Mo and $^{138}$Ba) and a strong
(in particular the lightest $p$-isotopes, $^{74}$Se, $^{76}$Kr, and $^{84}$Sr)
dependence on $Z$.

We discussed the still puzzling origin of $^{94}$Mo. Clearly, nuclear 
uncertainties cannot account for the factor of 10
deficiency in $^{94}$Mo abundance relative to other
$p$-abundances. The $^{94}$Mo production was found to depend on the
seeds in the Mo isotopes as well as on the density at which the
photodisintegration process occurs. This leaves room for possible
variations in the hydrodynamic history of the mainly contributing
explosive trajectories which could change the relative $p$-abundances.

By means of a simple Galactic chemical evolution code, including
$p$-process contributions at different metallicities, we explored the
SNIa contribution to the $p$-nuclei abundances in the Solar System.
We concluded that $p$-nuclei are mainly of primary-like origin, and
that SNIa can contribute at least 50\% to the solar abundance of all
$p$-nuclei provided that they result from standard Chandrasekhar-mass
delayed-detonations forming in the single-degenerate progenitor channel. 
Thus we identified a stellar source which, in
principle, is able to produce light and heavy $p$-nuclei almost at the
same level relative to $^{56}$Fe, including the much debated neutron magic
$^{92}$Mo and $^{96,98}$Ru. 

The important contribution from $p$-process nucleosynthesis to the
$s$-only nuclei $^{80}$Kr, $^{86}$Sr, and to the neutron magic
$^{90}$Zr has also been elaborated. Another relevant contribution is to
the neutron rich $^{96}$Zr, due to neutron captures from the residual
abundance of $^{22}$Ne during the explosive phase. With our Galactic
chemical evolution calculations, it was possible to predict a
significant contribution from SNIa in the considered scenario 
to the extinct $p$-radionuclides
$^{92}$Nb, $^{146}$Sm, and $^{96,98}$Tc in the early Solar System as
shown in Travaglio et al.~2014. Alternative scenarios for $p$-process
nucleosynthesis in binary systems have to be taken into account, such
as SNeIa from WD-WD mergers, where we also expect some $p$-process
production. A detailed analysis of different SNIa progenitors will be 
explored in a future work.

\acknowledgments This work has been supported by B2FH Association. The
numerical calculations have been also supported by Regione Lombardia
and CILEA Consortium through a LISA Initiative (Laboratory for
Interdisciplinary Advanced Simulation) 2010 grant. C.T. thanks
C. Arlandini, P. Dagna, and R. Casalegno for technical support in the
simulations.  C.T thanks the University of Washington (Institute for Nuclear Theory, 
Seattle, WA, INT Program INT-14-2b) for the kind support in the last phase of this work. 
The work of FKR was supported by the Deutsche
Forschungsgemeinschaft via the Emmy Noether Program (RO 3676/1-1) and
the graduate school ``Theoretical Astrophysics and Particle Physics''
at the University of W\"{u}rzburg (GRK 1147) Additional funding was
provided by the ARCHES prize of the German Ministry of Education and
Research (BMBF) and the DAAD/Go8 German-Australian exchange
program. TR is partially supported by the Swiss NSF, the European
Research Council (grant GA 321263-FISH). WH's work was supported by the
Transregional Collaborative Research Centre TRR33 ``The Dark
Universe'' und the Cluster of Excellence ``Origin and Structure of 
the Universe'' at Munich Technical University.

\clearpage

\begin{figure} 
\includegraphics[angle=-90,scale=.60]{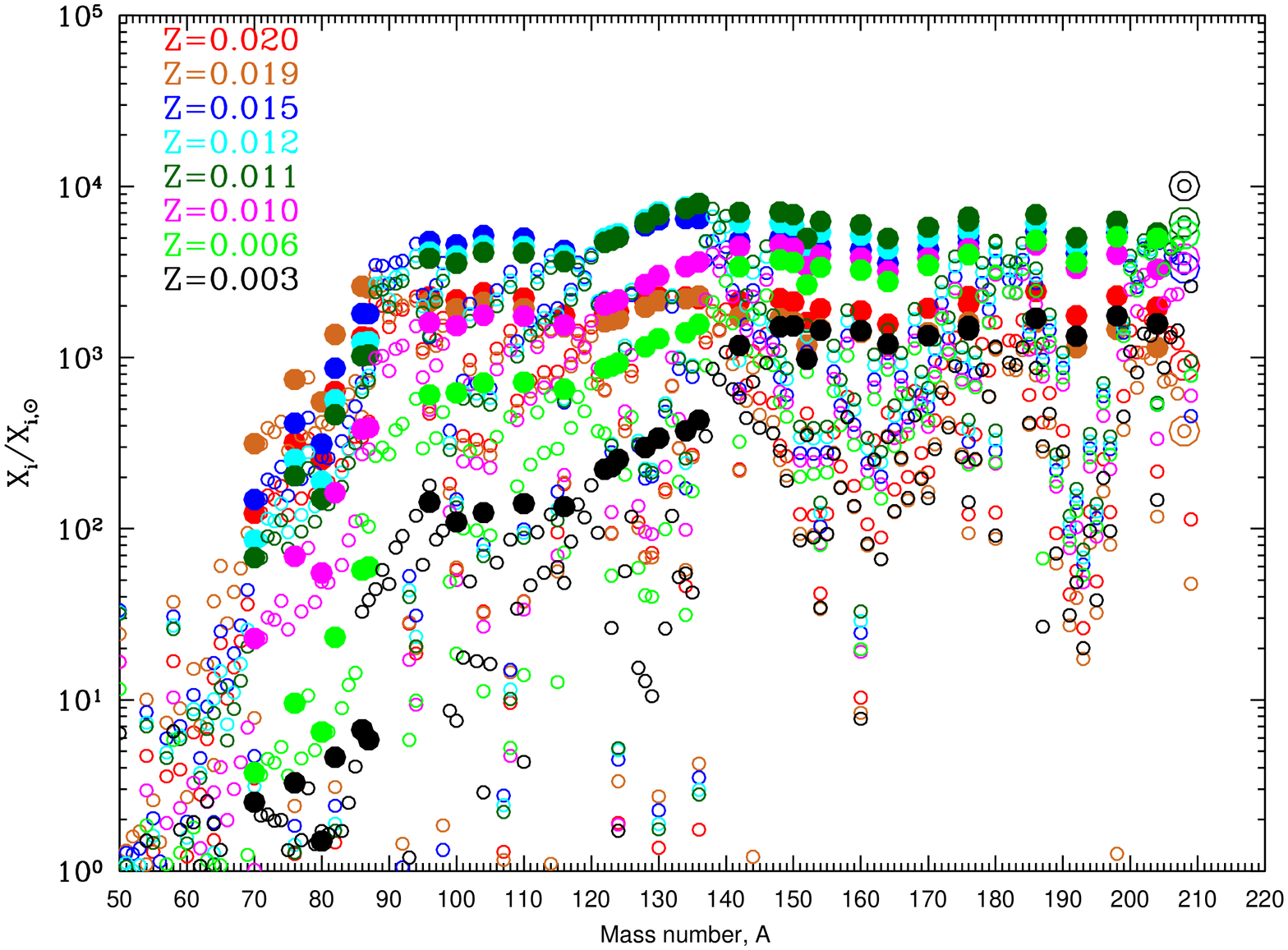} 
\caption{Different $s$-seed production factors relative to solar for the ST$\times$2 
$^{13}$C-pocket and $Z =$ 0.02 ({\it red}), $Z =$ 0.019 ({\it brown}), $Z =$ 0.015 ({\it blue}), $Z =$ 0.012 ({\it cyan}), $Z =$ 0.011 ({\it dark green}), 
$Z =$ 0.01 ({\it magenta}), $Z =$ 0.006 ({\it light green}), and $Z =$ 0.003 ({\it black}). Filled 
dots are for $s$-only isotopes. The {\it big open dot} is for $^{208}$Pb, see text for discussion. The solar values for 
this figure and all other figures of this paper are from Lodders~2009.}\label{fig1} \end{figure}

\begin{figure} 
\includegraphics[angle=-90,scale=.60]{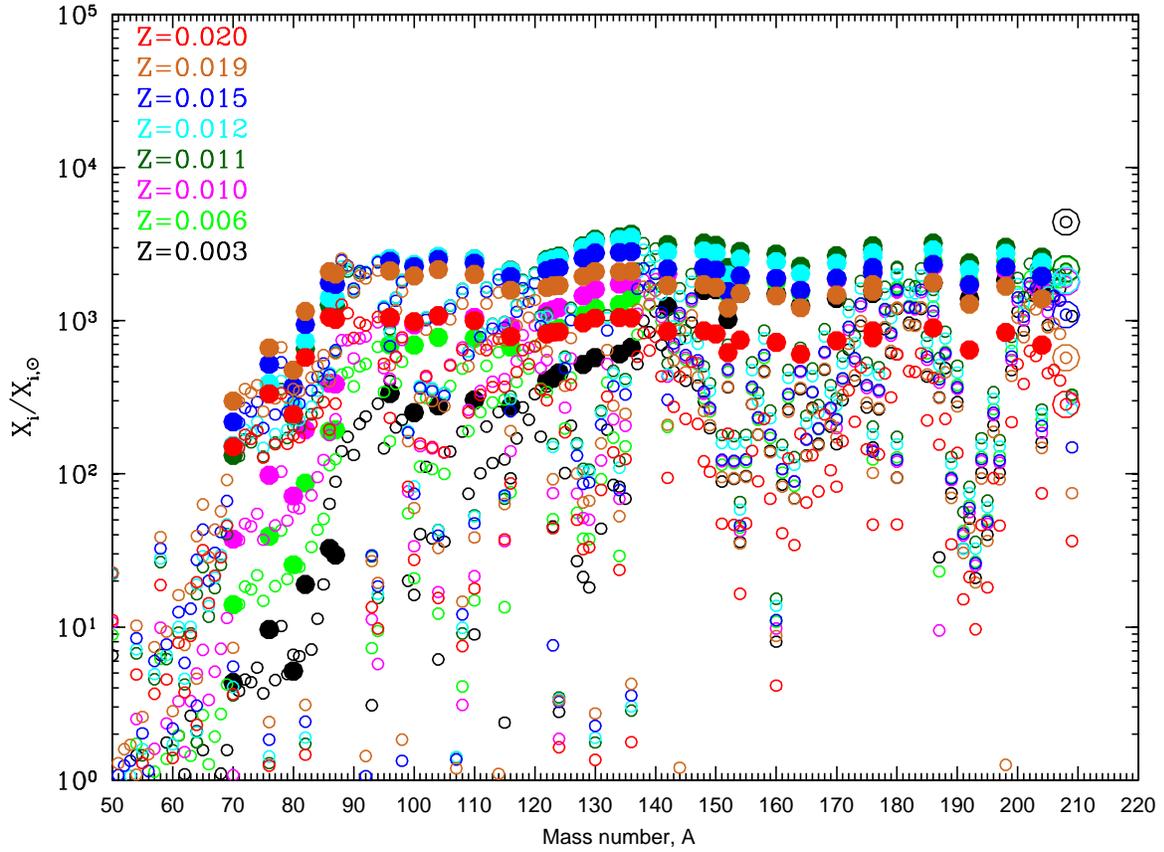} 
\caption{Different $s$-seed preduction factors relative to solar for an average of four 
different $^{13}$C-pockets (ST$\times$2, ST$\times$1.3, ST, ST/1.5 see text for discussion). Colours and symbols used are the 
same of Figure~1.}\label{fig2} \end{figure}

\begin{figure} 
\includegraphics[scale=.80]{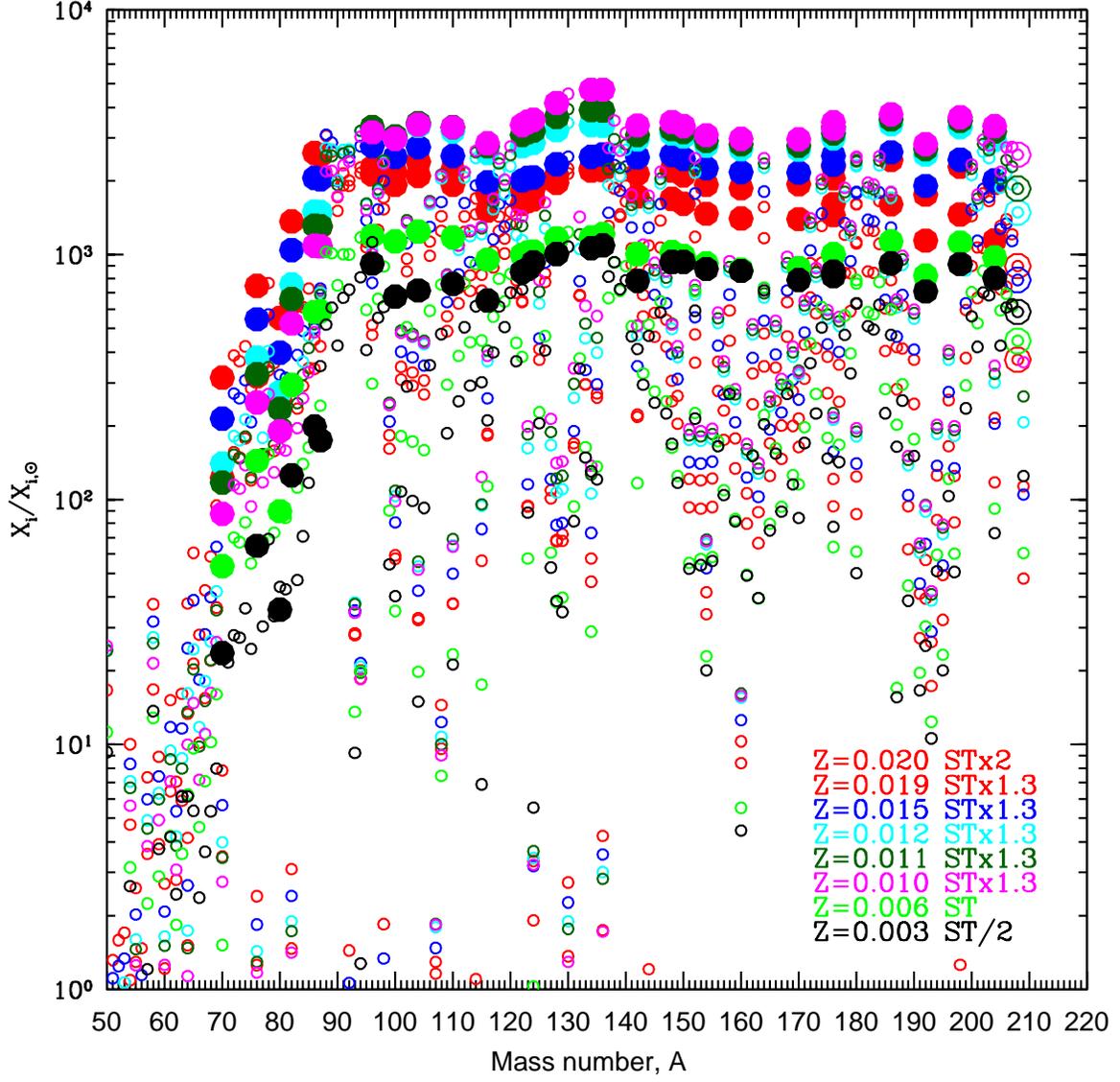} 
\caption{Distribution of $s$-seed abundances relative to solar for all
  $^{13}$C-pockets cases and metallicities covered, used for our Galactic chemical 
evolution calculation. Colours and symbols are the same of Figure~1. See text for a detailed discussion.}\label{fig3} \end{figure}

\begin{figure} 
\includegraphics[angle=-90,scale=.60]{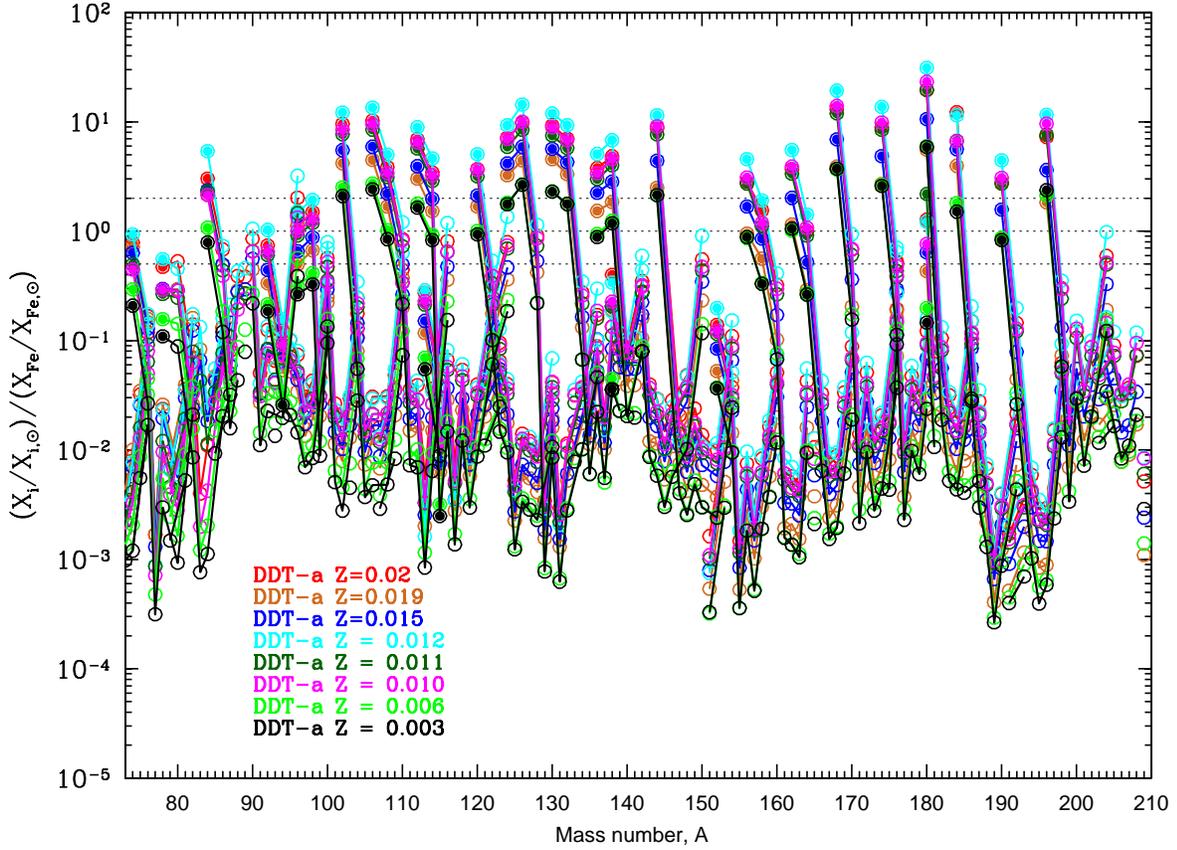} 
\caption{$p$-process yields normalized to solar and to Fe, obtained 
by using 51,200 tracer particles in the two-dimensional DDT-a model, different metallicities combined with their choosen $^{13}$C-pockets.
Colours are the same of Figure~1. Filled dots are for $p$-only isotopes.}\label{fig4} \end{figure}

\begin{figure} 
\includegraphics[angle=-90,scale=.60]{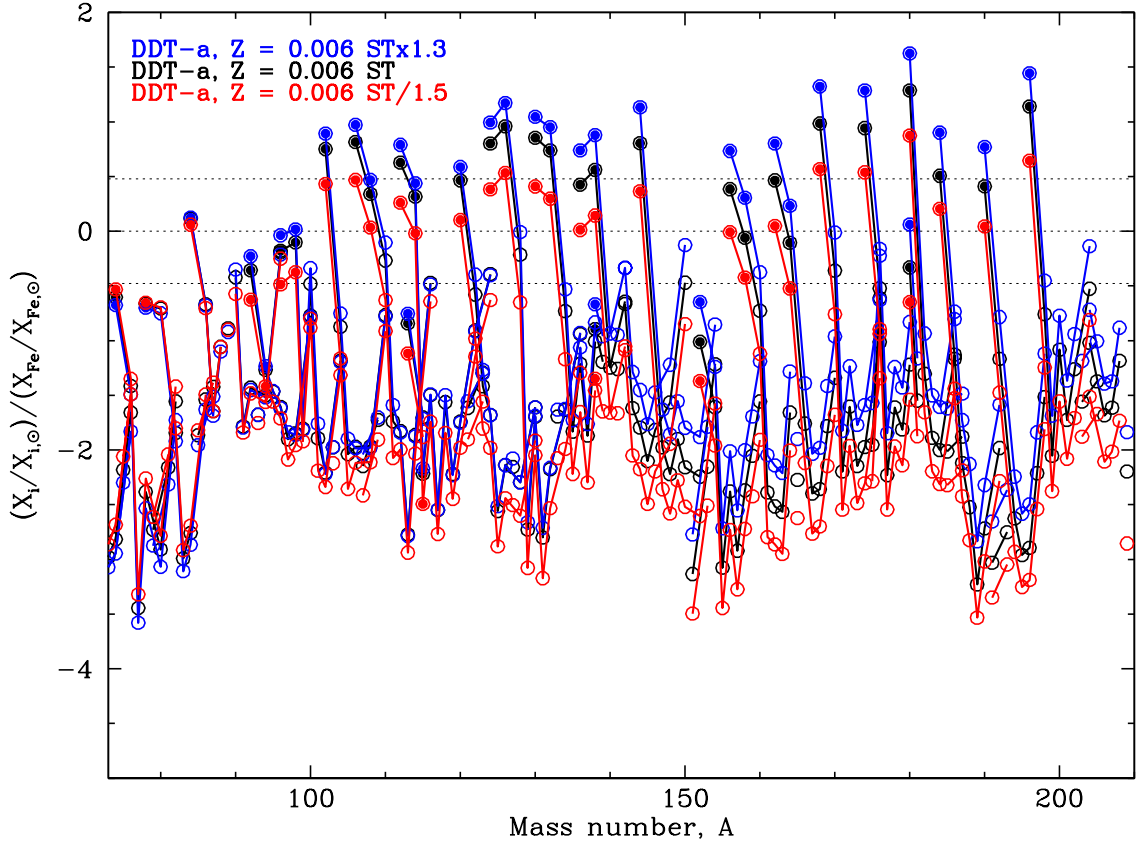} 
\caption{$p$-process yields normalized to solar and to Fe, obtained 
using 51,200 tracer particles in the two-dimensional DDT-a model, with a fixed metallicity ($Z =$0.006) and changing $^{13}$C-pocket 
abundance.}\label{fig5} \end{figure}

\begin{figure} 
\includegraphics[angle=-90,scale=.60]{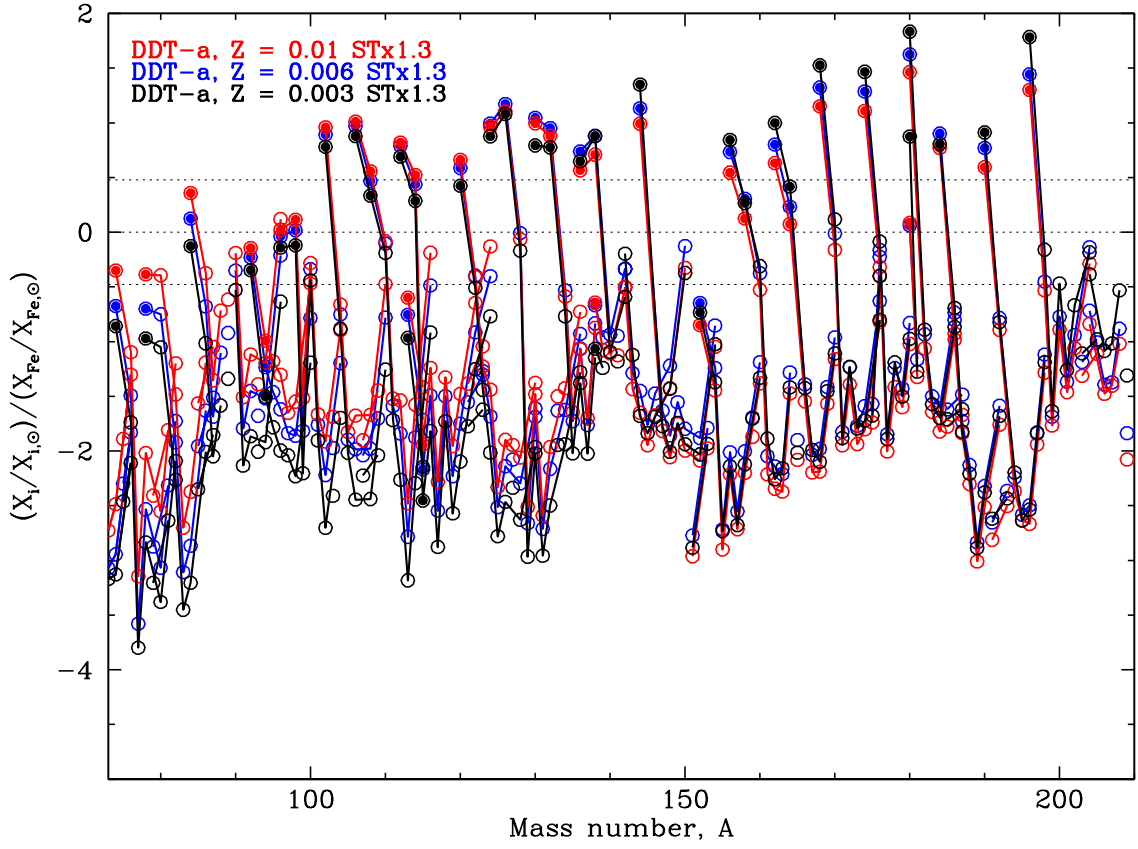} 
\caption{The same as Figure 5, but for this figure we fixed the $^{13}$C-pocket abundances (ST$\times$1.3) and changed the metallicity. 
}\label{fig6} \end{figure}

\begin{figure}
\includegraphics[angle=-90,scale=.60]{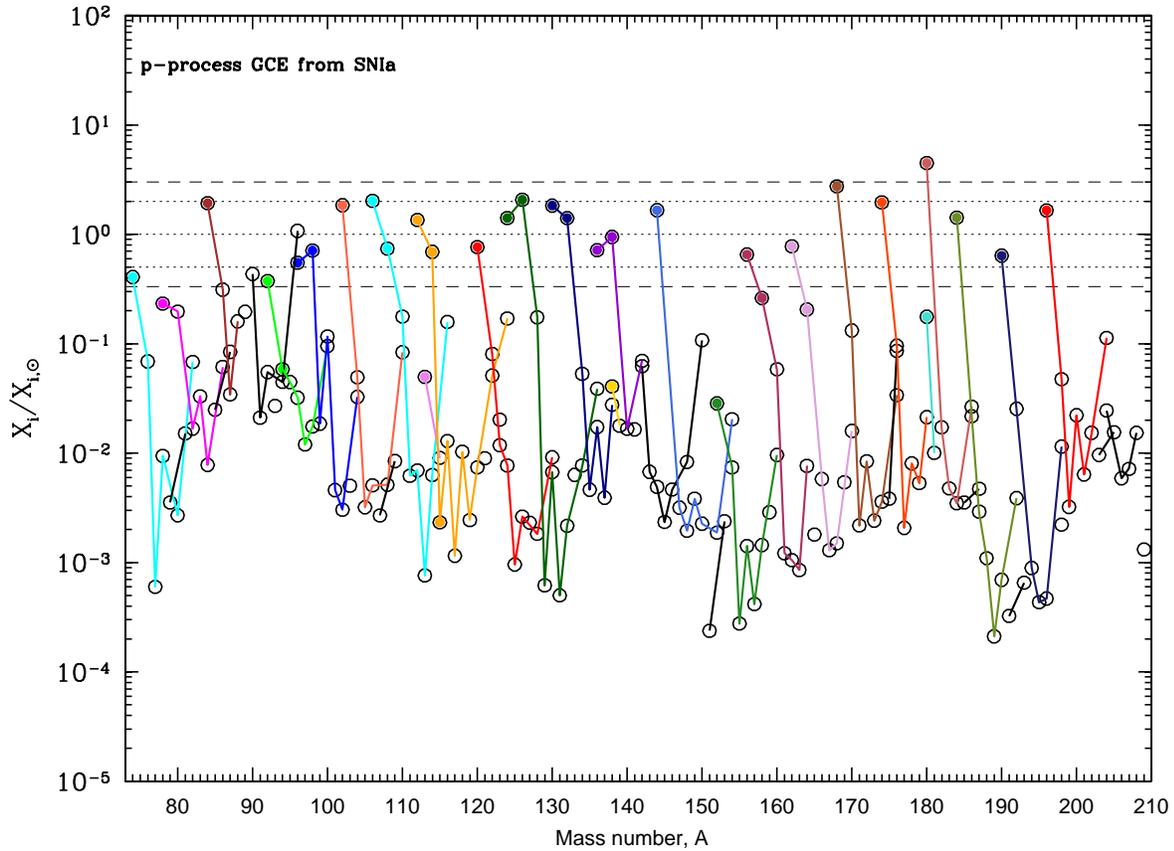}
\caption{Galactic chemical evolution of the $p$-process taken at the epoch of Solar System formation.
Filled dots are for the 35 isotopes classicaly defined as $p$-only. The isotopes of each element are connected by a line, and for each element
we adopt a different colour. For the $s$-seeds we used the abundances shown in Figure~3 and discussion in the text.}
\label{fig7}
\end{figure}

\begin{figure}
\includegraphics[width=\textwidth]{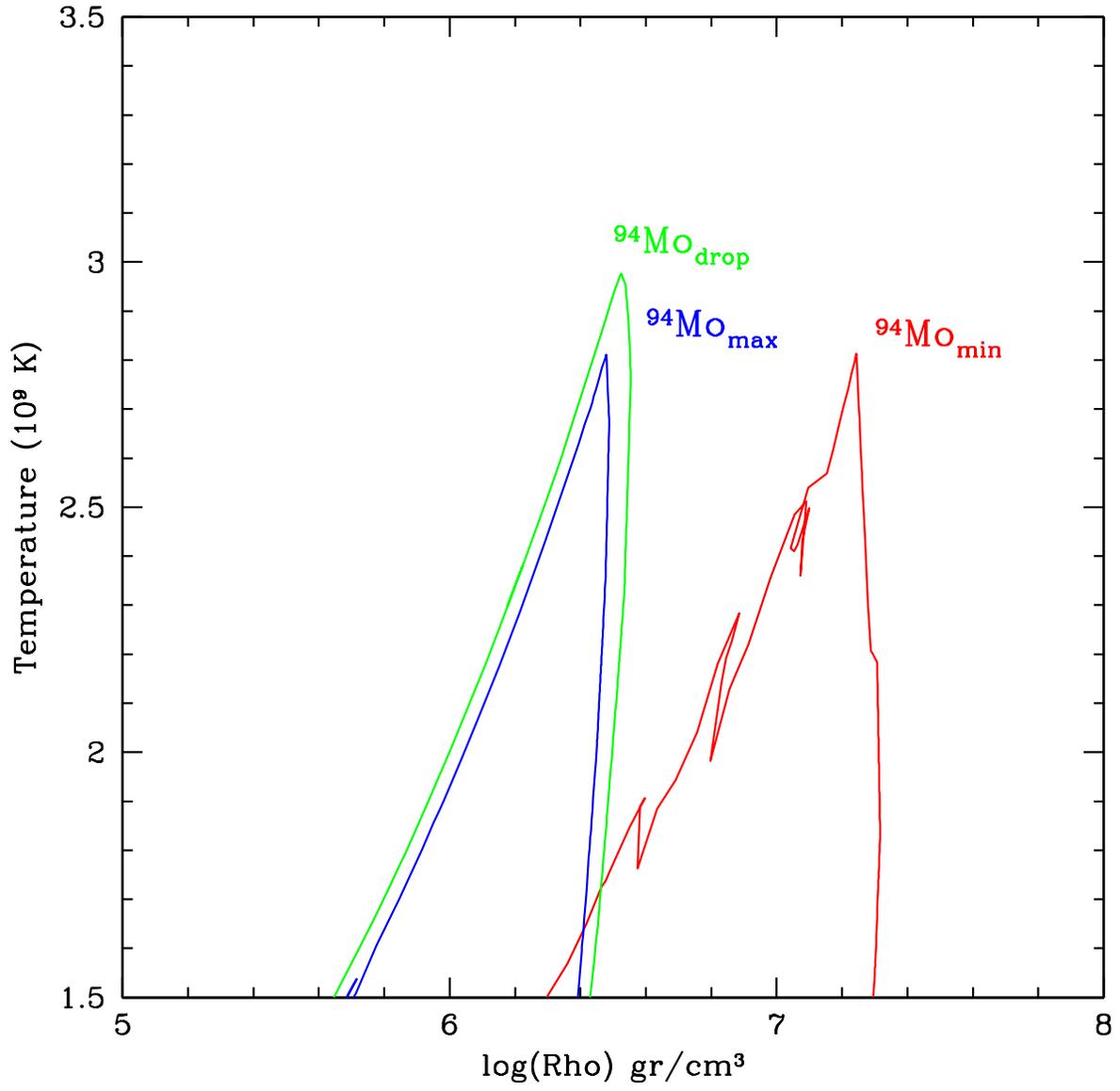}
\caption{\label{fig:flowTracers}Temperature vs density for three different tracers relevant for $^{94}$Mo production (see text for details).}
\label{fig8}
\end{figure}

\begin{figure}
\includegraphics[width=\textwidth]{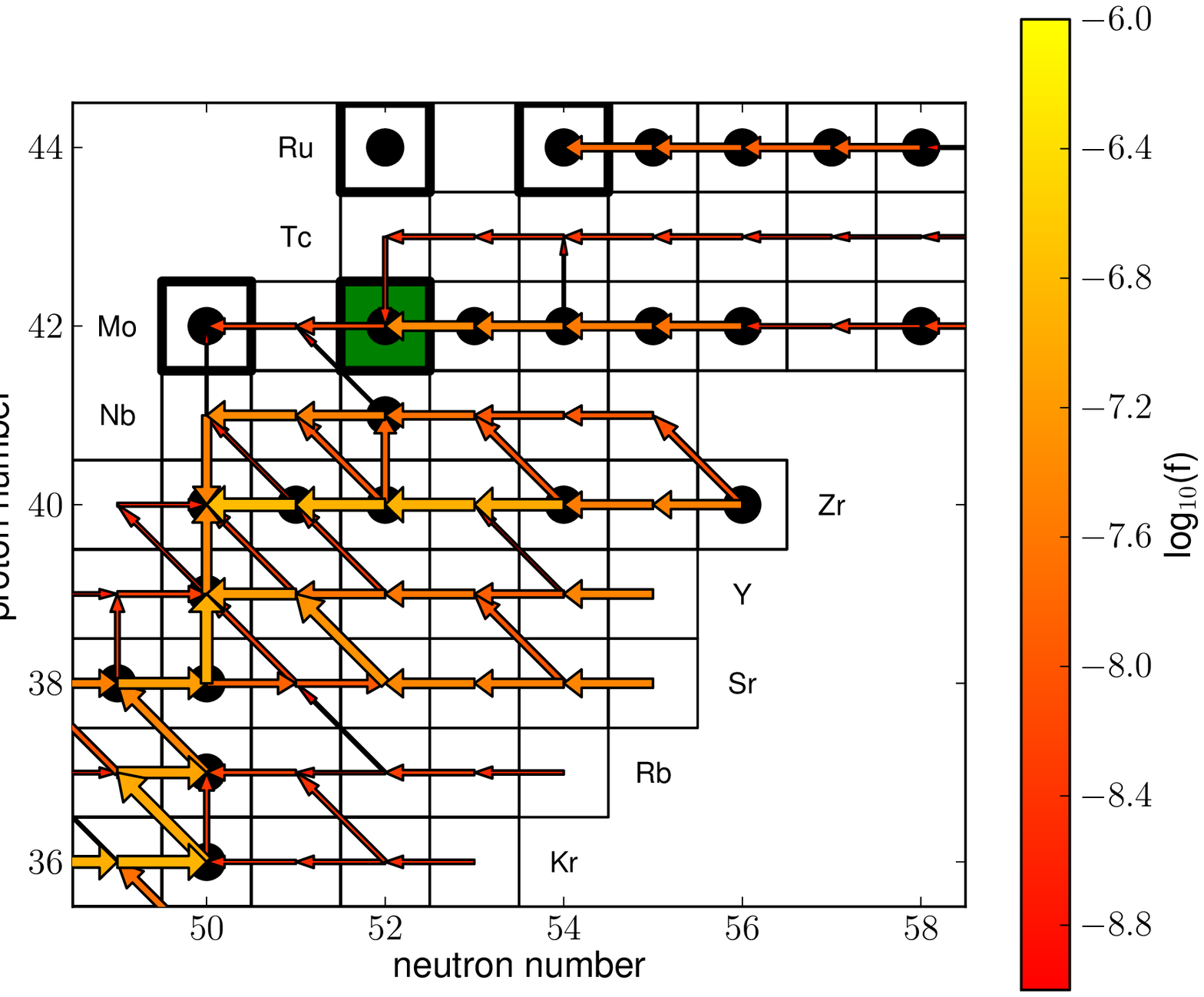}
\caption{\label{fig:flow94Momax}Reaction flow for maximum $^{94}$Mo production (trajectory $^{94}$Mo$_\mathrm{max}$ in Fig.\ \ref{fig:flowTracers}); size and color of the 
arrows relate to the magnitude of the time-integrated flux on a logarithmic scale. Only flows down to a factor 0.001 of the maximum flow are shown.}
\label{fig9}
\end{figure}

\begin{figure}
\includegraphics[width=\textwidth]{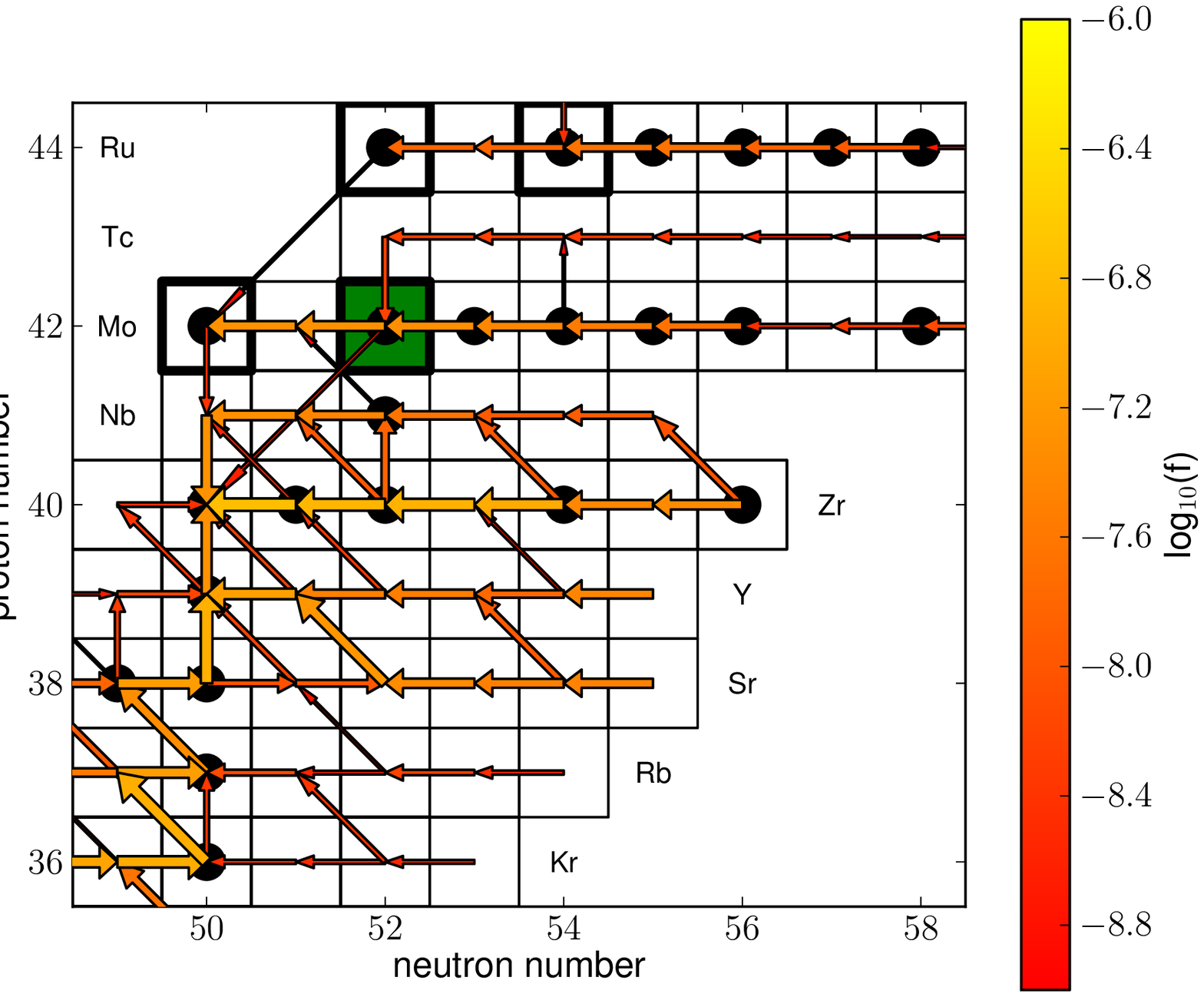}
\caption{\label{fig:flow94Modrop}Reaction flow for maximum $^{94}$Mo production beyond the dropping edge (trajectory $^{94}$Mo$_\mathrm{drop}$ in 
Fig.\ \ref{fig:flowTracers}); size and color of the arrows relate to the magnitude of the time-integrated flux on a logarithmic scale. Only flows down to a 
factor 0.001 of the maximum flow are shown.}
\label{fig10}
\end{figure}

\begin{figure}
\includegraphics[width=\textwidth]{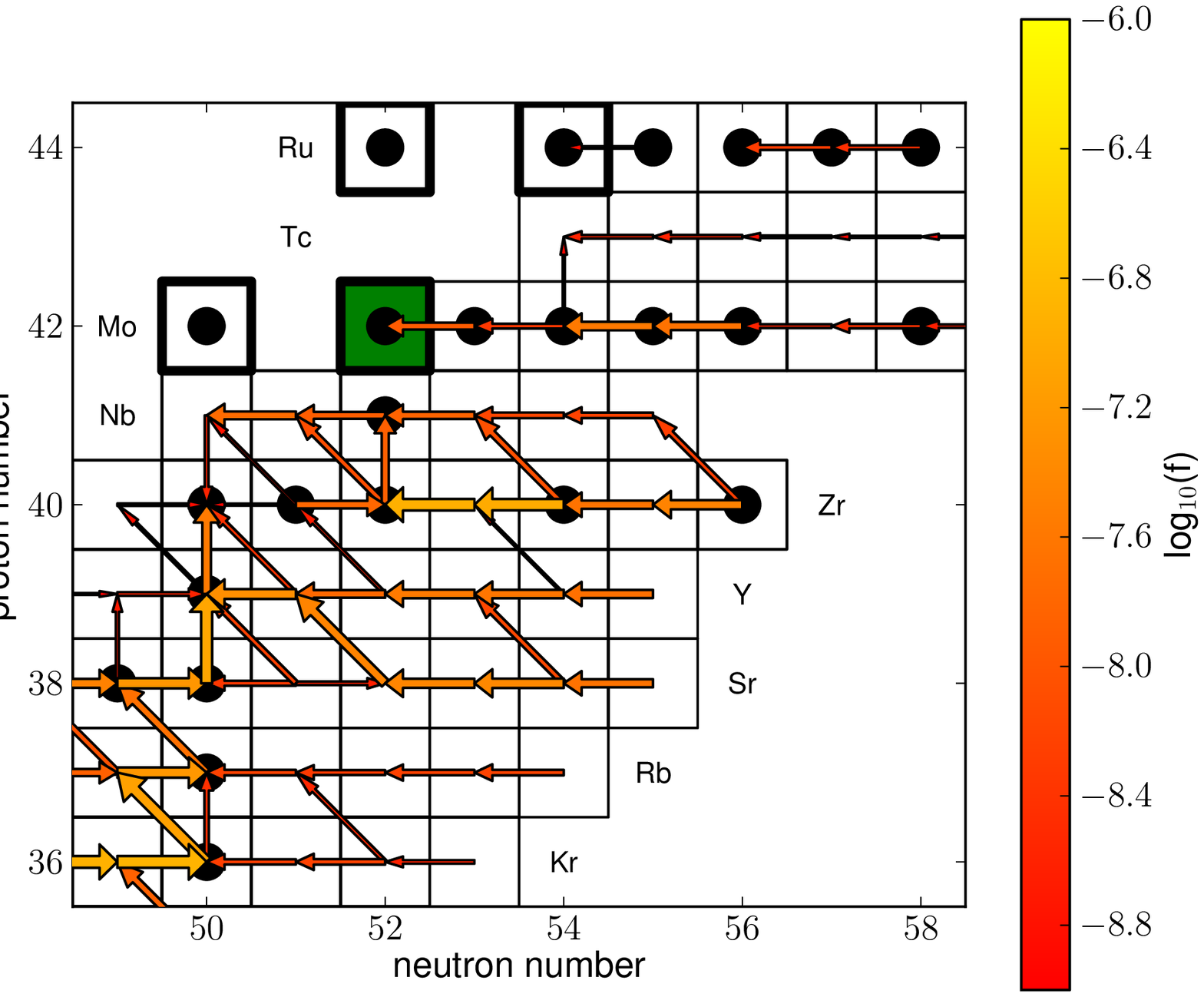}
\caption{\label{fig:flow94Momin}Reaction flow for a $^{94}$Mo production minimum (trajectory $^{94}$Mo$_\mathrm{min}$ in Fig.\ \ref{fig:flowTracers}); size and color of the arrows relate to the magnitude of the time-integrated flux on a logarithmic scale. Only flows down to a factor 0.001 of the maximum flow are shown.}
\label{fig11}
\end{figure}

\clearpage

\begin{deluxetable}{ccccccc}
\tabletypesize{\scriptsize}
\tablecaption{$p$-nuclides $Z =$ 0.006\label{tab:z0p006}}
\tablewidth{0pt}
\tablehead{
\colhead{Isotope} &  \colhead{STx2 $^{\rm (a)}$}  & \colhead{only $^{208}$Pb (STx2)} & \colhead{STx1.3 $^{\rm (a)}$} & \colhead{only $^{208}$Pb (STx1.3)} & 
\colhead{only heavy-$s$ (STx2)} & \colhead{only light-$s$ (STx2)} \\

\colhead{} &  \colhead{}  & \colhead{(\%)} & \colhead{} & \colhead{(\%)} & \colhead{(\%)} & \colhead{(\%)} \\

}

\startdata
$^{74}$Se  & 5.9766D-08 & 32 & 7.7692D-08 & 22 & 37 & 73 \\
$^{78}$Kr  & 2.2814D-08 & 14 & 2.6779D-08 & 5 & 22 & 60 \\
$^{84}$Sr  & 1.6684D-07 & 15 & 1.7652D-07 & 5 & 24 & 63 \\
$^{92}$Mo  & 3.3110D-07 & 20 & 2.5454D-07 & 10 & 33 & 46 \\
$^{94}$Mo  & 1.5926D-08 & 10 & 1.5816D-08 & 4 & 16 & 74 \\
$^{96}$Ru  & 1.5192D-07 & 22 & 1.1276D-07 & 11 & 35 & 44 \\
$^{98}$Ru  & 5.6763D-08 & 20 & 4.4197D-08 & 10 & 33 & 49 \\
$^{102}$Pd & 1.9007D-07 & 21 & 1.4257D-07 & 11 & 35 & 45 \\
$^{106}$Cd & 3.4325D-07 & 22 & 2.5156D-07 & 11 & 37 & 43 \\
$^{108}$Cd & 7.3755D-08 & 19 & 5.7087D-08 & 10 & 33 & 49 \\
$^{*113}$In & 2.3632D-09 & 14 & 1.9788D-09 & 6 & 24 & 59 \\
$^{112}$Sn & 4.4177D-07 & 21 & 3.2332D-07 & 11 & 36 & 43 \\
$^{114}$Sn & 1.2393D-07 & 17 & 9.9142D-08 & 8 & 30 & 54 \\
$^{*115}$Sn & 1.3516D-10 & 8 & 1.2630D-10 & 3 & 14 & 75 \\
$^{120}$Te & 3.0791D-08 & 15 & 2.5063D-08 & 7 & 28 & 56 \\
$^{124}$Xe & 1.1681D-07 & 18 & 8.7859D-08 & 9 & 36 & 45 \\
$^{126}$Xe & 1.6423D-07 & 20 & 1.1991D-07 & 10 & 40 & 40 \\
$^{130}$Ba & 9.3644D-08 & 6 & 8.6408D-08 & 2 & 14 & 79 \\
$^{132}$Ba & 8.2042D-08 & 12 & 6.7356D-08 & 6 & 32 & 56 \\
$^{*138}$La & 1.5568D-10 & - & 1.5268D-10 & - & - & 99 \\
$^{136}$Ce & 2.9490D-08 & 16 & 2.0296D-08 & 9 & 84 & - \\
$^{138}$Ce & 6.1838D-08 & 25 & 3.7265D-08 & 16 & 75 & - \\
$^{144}$Sm & 4.7213D-07 & 44 & 1.9701D-07 & 40 & 54 & - \\
$^{*152}$Gd & 4.2205D-10 & 6 & 2.8463D-10 & 3 & 94 & - \\
$^{156}$Dy & 4.5886D-09 & 29 & 2.3545D-09 & 22 & 71 & - \\
$^{158}$Dy & 2.4326D-09 & 11 & 1.5200D-09 & 7 & 89 & - \\
$^{162}$Er & 1.0113D-08 & 39 & 4.5391D-09 & 33 & 61 & - \\
$^{*164}$Er & 3.2957D-08 & 40 & 1.4267D-08 & 35 & 59 & - \\
$^{168}$Yb & 3.2922D-08 & 40 & 1.4330D-08 & 34 & 60 & - \\ 
$^{174}$Hf & 2.3801D-08 & 37 & 1.0562D-08 & 31 & 63 & - \\
$^{*180}$Ta & 1.0371D-11 & - & 6.4666D-12 & - & 100 & - \\
$^{180}$W  & 3.9411D-08 & 40 & 1.6578D-08 & 36 & 60 & - \\
$^{184}$Os & 3.8775D-09 & - & 2.2520D-09 & - & 100 & - \\
$^{190}$Pt & 5.2488D-09 & 27 & 2.3916D-09 & 23 & 73 & - \\
$^{196}$Hg & 9.9030D-08 & 58 & 3.5696D-08 & 61 & 41 & - \\ 

\enddata

\vspace{0.1em}
\hspace{9em}$^{(\rm a)}$ -- Nucleosynthesis yields.

\vspace{0.1em}
\hspace{9em}$^{(\rm *)}$ -- Isotopes pointed with * have to be excluded from the $p$-only list, as discussed by TRV11.

\vspace{0.1em}

\end{deluxetable}

\begin{deluxetable}{ccccccc}
\tabletypesize{\scriptsize}
\tablecaption{$s$-nuclides with important $p$-contribution, $Z =$0.006\label{tab:snuc}}
\tablewidth{0pt}
\tablehead{
\colhead{Isotope} &  \colhead{STx2$^{\rm (a)}$}  & \colhead{only $^{208}$Pb (STx2)} & \colhead{STx1.3$^{\rm (a)}$} & \colhead{only $^{208}$Pb (STx1.3)} &
\colhead{only heavy-$s$ (STx2)} & \colhead{only light-$s$ (STx2)} \\

\colhead{} &  \colhead{}  & \colhead{(\%)} & \colhead{} & \colhead{(\%)} & \colhead{(\%)} & \colhead{(\%)} \\

}

\startdata
$^{80}$Kr  & 1.2508D-07 & 12 & 1.5196D-07 & 4 & 19 & 58 \\
$^{86}$Sr  & 4.2708D-07 & 13 & 4.8676D-07 & 4 & 20 & 67 \\
$^{90}$Zr  & 3.2537D-06 & 17 & 2.8889D-06 & 7 & 28 & 55 \\
$^{96}$Zr  & 1.6979D-07 & - & 2.2666D-07 & - & - & 100 \\

\enddata

\vspace{0.1em}
\hspace{9em}$^{(\rm a)}$ -- Nucleosynthesis yields.

\vspace{0.1em}
\hspace{9em}$^{(\rm *)}$ -- Isotopes pointed with * have to be excluded from the $p$-only list, as discussed by TRV11.

\vspace{0.1em}

\end{deluxetable}

\clearpage

\begin{deluxetable}{ccc}
\tabletypesize{\scriptsize}
\tablecaption{Galactic chemical evolution of $p$-nuclides\label{tab:gce}}
\tablewidth{0pt}
\tablehead{
\colhead{Isotope} &  \colhead{GCE} &  \colhead{(GCE/Solar)} \\

}

\startdata
$^{74}$Se  & 4.186D-10 & 0.41 \\
$^{78}$Kr  & 9.968D-11 & 0.23 \\
$^{84}$Sr  & 5.760D-10 & 1.93 \\
$^{92}$Mo  & 3.643D-10 & 0.37 \\
$^{94}$Mo  & 3.663D-11 & 0.06 \\
$^{96}$Ru  & 1.431D-10 & 0.55 \\
$^{98}$Ru  & 6.294D-11 & 0.71 \\
$^{102}$Pd & 7.251D-11 & 1.84 \\
$^{106}$Cd & 1.173D-10 & 2.02 \\
$^{108}$Cd & 3.067D-11 & 0.74 \\
$^{*113}$In & 1.232D-12 & 0.05 \\
$^{112}$Sn & 1.451D-10 &  1.35 \\
$^{114}$Sn & 5.177D-11 &  0.69 \\
$^{*115}$Sn & 8.794D-14 &  0.002 \\
$^{120}$Te & 1.254D-11 &  0.76 \\
$^{124}$Xe & 3.350D-11 &  1.41 \\
$^{126}$Xe & 4.281D-11 &  2.07 \\
$^{130}$Ba & 3.186D-11 &  1.83 \\
$^{132}$Ba & 2.364D-11 &  1.40 \\
$^{*138}$La & 5.825D-14 &  0.04 \\
$^{136}$Ce & 5.816D-12 &  0.72 \\
$^{138}$Ce & 1.045D-11 &  0.95 \\
$^{144}$Sm & 5.238D-11 &  1.66 \\
$^{*152}$Gd & 8.258D-14 & 0.03  \\
$^{156}$Dy & 6.448D-13 &  0.65 \\
$^{158}$Dy & 4.439D-13 &  0.26 \\
$^{162}$Er & 1.231D-12 &  0.78 \\
$^{*164}$Er & 3.792D-12 &  0.20 \\
$^{168}$Yb & 3.792D-12 &  2.75 \\ 
$^{174}$Hf & 2.802D-12 &  1.96 \\
$^{*180}$Ta & 2.261D-15 &  0.18 \\
$^{180}$W  & 4.416D-12 &  4.48 \\
$^{184}$Os & 7.157D-13 &  1.42 \\
$^{190}$Pt & 6.640D-13 &  0.64 \\
$^{196}$Hg & 8.884D-12 &  1.65 \\ 

\enddata

\vspace{0.1em}
\hspace{9em}$^{(\rm *)}$ -- Isotopes pointed with * have to be excluded from the $p$-only list, as discussed by TRV11.

\vspace{0.1em}

\end{deluxetable}

\begin{deluxetable}{ccc}
\tabletypesize{\scriptsize}
\tablecaption{Galactic chemical evolution of $s$-nuclides with important $p$-contribution \label{tab:sgce}}
\tablewidth{0pt}
\tablehead{

\colhead{Isotope} &  \colhead{GCE}  & \colhead{(GCE/solar)} \\

}

\startdata
$^{80}$Kr  & 5.610D-10 & 0.20 \\
$^{86}$Sr  & 1.692D-09 & 0.31 \\
$^{90}$Zr  & 5.734D-09 & 0.43 \\
$^{96}$Zr  & 8.258D-10 & 1.08 \\

\enddata

\vspace{0.1em}
\hspace{9em}$^{(\rm *)}$ -- Isotopes pointed with * have to be excluded from the $p$-only list, as discussed by TRV11.

\vspace{0.1em}

\end{deluxetable}

\end{document}